\documentclass[floatfix,preprint,prl,aps,showpacs]{revtex4-1}
\usepackage{graphicx}
\def\e{\varepsilon}

\begin{document}

\title{Quantitative analysis of valence photoemission spectra 
and quasiparticle excitations \\ at chromophore/semiconductor interfaces}

\author{Christopher E. Patrick}
\author{Feliciano Giustino}

\affiliation{Department of Materials, University of Oxford, Parks Road,
Oxford OX1 3PH, United Kingdom}

\begin{abstract}
Investigating quasiparticle excitations of molecules on surfaces through photoemission spectroscopy
forms a major part of nanotechnology research. Resolving spectral features at these interfaces 
requires a comprehensive theory of electron removal and addition processes in molecules and solids 
which captures the complex interplay of image charges, thermal effects and configurational disorder.
We here develop such a theory and calculate the quasiparticle energy-level alignment and the valence 
photoemission spectrum for the prototype biomimetic solar cell interface between anatase TiO$_2$ and 
the N3 chromophore. By directly matching our calculated photoemission spectrum to experimental
data we clarify the atomistic origin of the chromophore peak at low binding energy. This case study 
sets a new standard in the interpretation of photoemission spectroscopy at complex chromophore/semiconductor 
interfaces. 
\end{abstract}
\date{\today}
\maketitle

The interfaces between solids and molecules form the backbone of many areas of nanotechnology 
research~\cite{Somorjai2011}, including biomimetic
photovoltaics~\cite{Gratzel20012}, molecular electronics~\cite{Nitzan2003}, 
surface-transfer doping~\cite{Wehling2008}, catalysis~\cite{Marshall2010}, and 
photocatalysis~\cite{Fujishima2008}. 
Developing an understanding of electron energetics and dynamics at these interfaces
is an essential step towards rational strategies of
device optimization, as well as a key challenge for the theory
of excitations in highly anisotropic and inhomogeneous systems.

For example, in dye-sensitized solar cells electrical current is generated 
by electron injection from a photo-excited molecular chromophore into a wide gap 
semiconductor~\cite{Gratzel20012}. This process crucially relies on the 
type-II alignment between the discrete energy levels of the chromophore and the 
energy bands of the semiconductor. In this context a quantitative understanding of 
the energetics of charge transfer across the interface, as provided by the quasiparticle 
spectrum, is important for gaining insight into the physics of charge injection.

The prototypical dye-sensitized solar cell interface consists of N3 dyes 
[Ru(dcbpyH$_2$)$_2$(NCS)$_2$] adsorbed on anatase TiO$_2$~\cite{Nazeeruddin1993}.
\begin{figure}
\includegraphics[width=0.7\columnwidth]{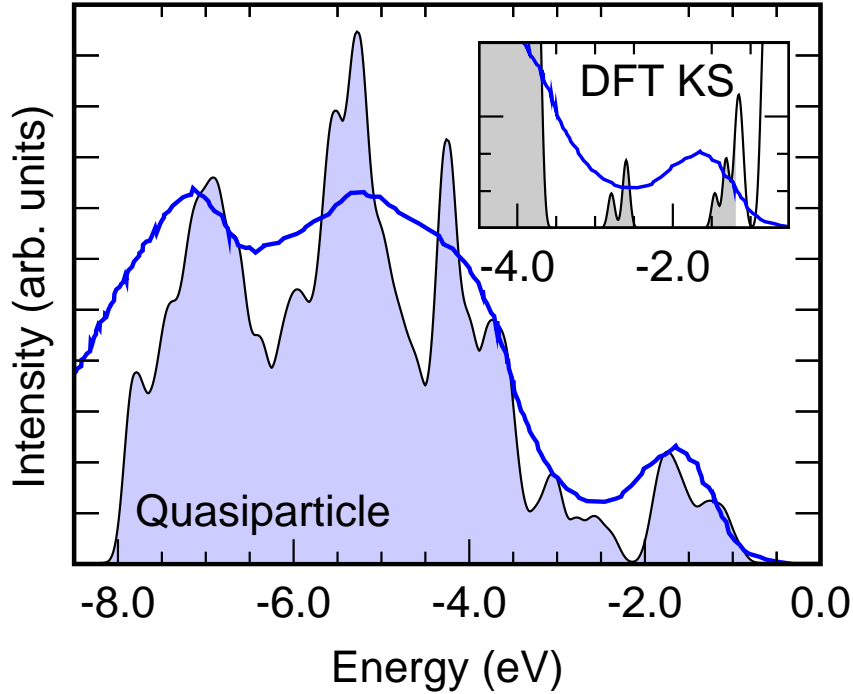}
\caption{
\label{fig.1} (Color online) 
Valence photoemission spectrum of the N3/TiO$_2$(101) interface:
measured PES spectrum of Ref.~\citenum{Hahlin2011} [blue (grey) line] and
our quasiparticle calculation using Eqs.~(\ref{eq.1}) and (\ref{eq.2}) (black line
and light blue shaded area).
Inset: comparison between the measured PES spectrum of Ref.~\citenum{Hahlin2011}
[blue (grey) line] and the KS density of states at the N3/TiO$_2$ interface (black
line and grey shaded area).
In the DFT KS calculation there is no gap between occupied (shaded regions) and 
unoccupied (unshaded) states.
}
\end{figure}
The quasiparticle spectrum of this interface, as measured by valence photoemission 
spectroscopy (PES) experiments, has been reported in several
studies~\cite{Hahlin2011, Rensmo1997,Hagfeldt2000, Schwanitz2007, Westermark20023}.
The most recent measurements~\cite{Hahlin2011} are shown in Fig.~\ref{fig.1} 
(blue line).

The standard qualitative interpretation of the N3/TiO$_2$ PES spectrum is to
assign the feature at a binding energy of $\sim$1.8~eV to the highest 
occupied molecular orbital (HOMO) of the N3 dye, and the leading edge
at a binding energy of $\sim$3~eV to the valence band top (VBT) of TiO$_2$.
However, owing to considerable spectral broadening in the range of 1-2~eV, 
the precise location of the TiO$_2$ VBT in the spectrum is unkown 
and the unanbiguous assignment of the feature at low-binding energy
to specific excitations in the dye is still missing. In this context
a first-principles calculation of the PES spectrum is highly desirable 
as it would enable a {\it quantitative} interpretation of the measured 
spectrum based on the underlying interface structure at the atomic scale.
Despite considerable progress in this area~\cite{Neaton2006,Angelis2007b, Marom2011}, 
systematic comparisons of calculated energy-level alignments and quasiparticle 
spectra with PES have been scarce so far. 
This situation is to be ascribed to the inherent complexity of 
chromophore/semiconductor interfaces, and to the lack of quasiparticle 
techniques suitable for atomistic models with hundreds of atoms.

In this work we propose a procedure for calculating from first principles 
quasiparticle energy-level alignments and valence photoemission spectra
at complex chromophore/semiconductor interfaces. Our procedure combines 
density-functional calculations on large interface models, bulk GW calculations, 
molecular $\Delta$SCF calculations, image-charge renormalization, thermal broadening, 
and configurational disorder to obtain the most accurate possible quasiparticle 
spectrum at the interface.

Within the sudden approximation, in a PES experiment the photocurrent for a
binding energy $\e$ is proportional to the quasiparticle spectral function 
$A(\e)$ through a factor $I_0$ incorporating matrix elements and escape-depth
effects~\cite{Hufnerbook,Hedin1965,Damascelli2003}. At small binding energy quasiparticles are 
well defined excitations and the spectral function is given by the quasiparticle 
density of states. Under these standard assumptions the calculation of valence 
PES spectra reduces to the calculation of quasiparticle energies 
for occupied electronic states.

\begin{figure}
\includegraphics[width=\columnwidth]{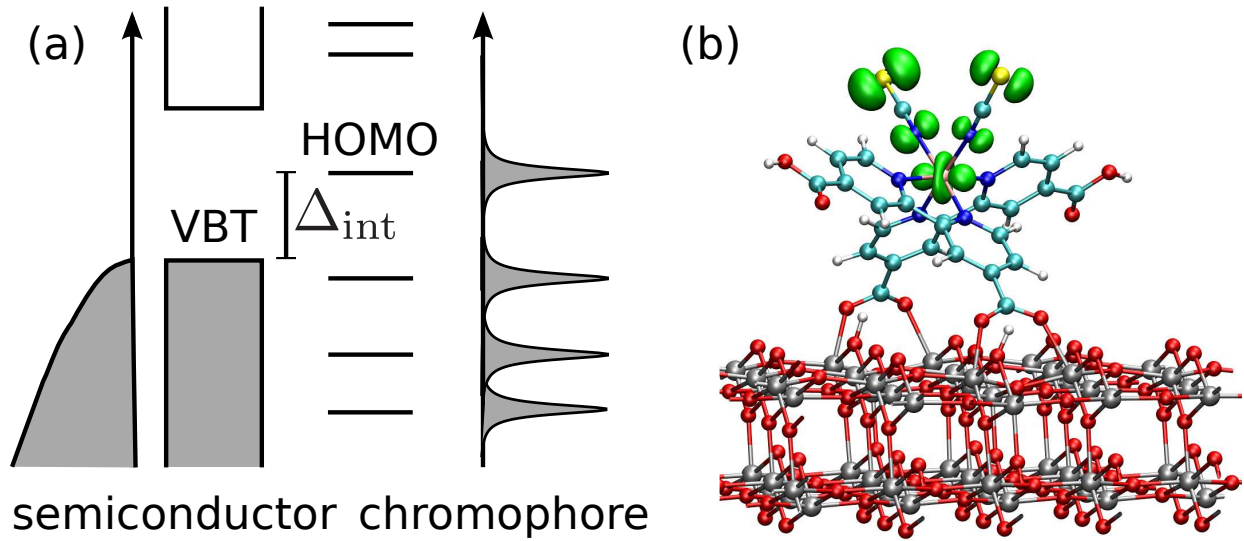}
\caption{\label{fig.2} (Color online)
(a) Schematic representation of the calculation procedure corresponding to
Eq.~(\ref{eq.1}): the spectral functions of semiconductor and chromophore
obtained from separate calculations are superimposed after calculating the
HOMO/VBT offset $\Delta_{\rm int}$ through Eq.~(\ref{eq.2}).
(b) Atomistic model of a representative N3/TiO$_2$(101) interface
[model I2b of Ref.~\citenum{Patrick2011}]
and isodensity plot of the highest occupied KS state, i.e. the N3 HOMO.
The isodensity is 0.03~\AA$^{-3}$, and the atom color code is:
Ti (silver), O (red), C (cyan), N (blue), H (white), Ru (pink).
}
\end{figure}

As GW quasiparticle calculations at interfaces~\cite{Neaton2006,Rignanese2008} 
are computationally prohibitive for large systems such as the N3/TiO$_2$ interface 
considered here (299 atoms), we need to devise a practical alternative retaining 
the accuracy of a complete GW calculation. For this purpose we partition 
the spectral function of the interface into contributions from the bulk semiconductor
$A_{\rm s}$ and the isolated chromophore $A_{\rm c}$, and evaluate 
these contributions separately. 
If $A_{\rm s}$ and $A_{\rm c}$ are referred to the
VBT
of the semiconductor and to the 
HOMO of the molecule, respectively, 
and $\Delta_{\rm int}$ 
indicates the difference between the quasiparticle energies of VBT and HOMO 
at the interface, then the photocurrent is obtained as 
  \begin{equation}\label{eq.1}
  I(\e)=I_{0,\rm s} A_{\rm s}(\e)+I_{0,\rm c} A_{\rm c}(\e+\Delta_{\rm int}).
  \end{equation} 
This simple procedure is illustrated in Fig.~\ref{fig.2}(a). 

The partitioning that we propose is legitimate as long as the frontier orbitals 
of the molecule do not hybridize with the valence manifold of the oxide,
i.e. they are well separated from the substrate and retain the original gas-phase 
character. This condition is verified in most type-II interfaces between large
chromophores and wide-gap semiconductors~\cite{Martsinovich2010}, and is also common in the case of molecular 
physisorption \cite{Kanai2010}. 

We now proceed to describe how $\Delta_{\rm int}$, $A_{\rm s}$, $A_{\rm c}$,
$I_{0,\rm s}$, and $I_{0,\rm c}$ are calculated. The HOMO/VBT offset 
$\Delta_{\rm int}$ is obtained as:
  \begin{equation}
  \Delta_{\rm int} =   (\e_{\rm H}^{\rm KS} + \Delta \e^{\rm img}_{\rm H} +
  \Delta \e^{\rm QP}_{\rm H} )
  - (\e_{\rm V}^{\rm KS} + \Delta\e^{\rm slab}_{\rm V} + 
  \Delta \e^{\rm QP}_{\rm V}).
  \label{eq.2}
  \end{equation}
In this equation $\e_{\rm V}^{\rm KS}$ and $\e_{\rm H}^{\rm KS}$ are 
the standard density-functional theory (DFT) Kohn-Sham (KS) eigenvalues of the semiconductor
and of the molecule, respectively, both obtained from an interface calculation.
$\Delta\e^{\rm slab}_{\rm V}$ is a correction to the valence band top of the 
semiconductor which removes spurious quantum confinement and surface effects 
associated with the semiconductor slab in the interface model~\cite{VandeWalle1987}. 
$\Delta \e^{QP}_{\rm V}$ is the GW quasiparticle shift of the semiconductor VBT,
and is obtained from a separate bulk calculation. $\Delta \e^{\rm QP}_{\rm H}$ 
is the quasiparticle renormalization of the molecular HOMO. This is determined 
by explicitly computing electron removal energies for the isolated molecule 
using the $\Delta$SCF method. $\Delta \e^{\rm img}_{\rm H}$ is 
the substrate-induced renormalization of the molecular HOMO energy, and arises 
from the dielectric screening by the semiconductor of the photo-hole in the 
chromophore~\cite{Neaton2006}.
In addition to reducing the complexity of the calculation, the decomposition
of $\Delta_{\rm int}$ in Eq.~(\ref{eq.2}) allows us to determine
$\Delta \e^{QP}_{\rm V}$ and $\Delta \e^{\rm QP}_{\rm H}$ using the most 
accurate techniques available for bulk solids and molecules, respectively. 
This point 
is especially relevant since 
molecular systems require going beyond many-body 
GW perturbation theory using self-consistency or improved starting 
Hamiltonians~\cite{Blase2011, Rostgaard2010, Stenuit2010, Stan2006}.

The spectral function of the chromophore $A_{\rm c}(\e)$ with the zero of
energy set at the HOMO is calculated using the DFT KS density of states,
broadened to include electron-vibration interactions.
This choice corresponds to applying the same quasiparticle shift
to all the molecular levels of the dye, an approximation which is expected
to hold quite generally for states near the HOMO~\cite{Stenuit2010}. 
The broadening of $A_{\rm c}(\e)$ is calculated 
within the adiabatic approximation \cite{Grimvall1981} by performing explicit first-principles 
molecular dynamics simulations at the experimental temperature.
The spectral function of the semiconductor substrate $A_{\rm s}(\e)$
is obtained in the same way as for the chromophore. 
Lifetime broadening arising 
from electron-electron interactions is vanishing below the threshold for 
electron-hole pair generation (i.e. the semiconductor band gap) hence is not 
explicitly considered in our description.
Configurational disorder is included 
in Eq.~(\ref{eq.1}) through $\Delta_{\rm int}$ by performing 
separate calculations on the most likely interface conformations.

The prefactors $I_{0,\rm s}$ and $I_{0,\rm c}$ in Eq.~(\ref{eq.1}) depend on
surface coverage, inelastic scattering of the photoelectrons, transmission
losses, and dipole matrix elements. 
The evaluation of these terms still
poses a significant challenge~\cite{Stojic2008, Husser2011}.
For simplicity it is
convenient to assume that, at low binding energy, the size of the dipole 
matrix elements in the semiconductor and the molecule is similar.
This approximation is sensible for organic and metal-organic chromophores 
adsorbed on semiconducting oxides where both VBT and HOMO
have significant $p$-character \cite{Greiner2012}. 
Dissipative effects 
are taken into account using a phenomenological escape-depth model based on 
the universal curve of inelastic mean free paths~\cite{Hufnerbook}, 
and surface coverage is chosen 
so as to match the experimental conditions. With these choices the
energy dependence of the prefactors drops, and their ratio $I_{0,\rm s}/I_{0,\rm c}$ 
sets the relative intensity of the photocurrents originating 
in the substrate and in the molecular layer.

After illustrating our theoretical framework we now describe the computational
details. All calculations are performed within the generalized gradient approximation 
to DFT of Ref.~\citenum{Perdew1996}, using ultrasoft pseudopotentials~\cite{Vanderbilt1990}
and planewaves basis sets as implemented in the \texttt{Quantum ESPRESSO}  
distribution~\cite{quantumespresso}. The Ti semicore 3$s$ and 3$p$ states are 
explicitly included and the kinetic energy cutoffs for wavefunctions and 
charge density are 35 and 200~Ry, respectively. The atomistic structures of 
the TiO$_2$/N3 interface models are described in detail in our previous work 
Ref.~\citenum{Patrick2011}.
In these models the N3 dye is anchored to the Ti atoms of the anatase TiO$_2$(101) 
surface through the O atoms of its carboxylic groups [Fig.~\ref{fig.2}(b)].
We consider six possible 
adsorption geometries in order to cover all the interface models studied 
in the literature \cite{Nazeeruddin2003,Rensmo1999,Schiffmann2010,Angelis2010}. 
The typical size of our extended interface models 
is of 299 atoms, and for all models we use $\Gamma$-point sampling.
In order to prevent incorrect orbital occupancies at the N3/TiO$_2$ interface
we include a Hubbard $U$ correction~\cite{Ansimov1991} for the Ti-3$d$ states,
using the method of Ref.~\citenum{Cococcioni2005}. The Hubbard $U$ parameter
is set to 7.5~eV, as determined from first principles using a self-consistent
$GW$+$U$ calculation~\cite{Patrick2012}.
The energy levels of the TiO$_2$ slab are corrected for finite-size effects
by referring all the excitation energies to the Ti-3$s$ semicore states.
The quasiparticle corrections for bulk anatase TiO$_2$ are calculated within
the $G_0W_0$ approximation~\cite{Hedin1965, Hybertsen1986, Onida2002} using
the self-consistent $GW$+$U$ procedure described in Ref.~\citenum{Patrick2012}.

The ionization potential (IP) of the chromophore is determined using the $\Delta$SCF
method. The spurious Coulomb interaction between periodic replicas is eliminated
using the Coulomb truncation technique of Ref.~\citenum{Martyna1999}.
We ran extensive test calculations on molecules ranging from thiophene to fullerenes, and
obtained IPs with a mean average error with respect to experiment of 0.22~eV.
This level of accuracy is similar to those reported in 
Refs.~\citenum{Blase2011,Rostgaard2010} for similar molecules, and superior 
to the best GW calculations reported so far, which yield errors 
of 0.32~eV~\cite{Blase2011} and 0.4~eV~\cite{Rostgaard2010} on average.
In order to determine both the HOMO energy of the chromophore and the
image-charge renormalization from the substrate we perform two separate calculations.
In the first calculation we determine the removal energy of one electron 
in the isolated chromophore.
In the second calculation we determine the removal energy of one electron 
(from the dye) in a finite model of the chomophore/semiconductor 
interface~\cite{Angelis2007, Lundqvist22006}. The difference of these 
two calculations yields by construction the image-charge renormalization of the HOMO.
The nanocluster model contains 68 TiO$_2$ units, and is obtained by cutting 
a slab of bulk anatase along the (101) direction.

The broadening of energy levels due to finite temperature is calculated
using Car-Parrinello molecular dynamics~\cite{CarParrinello,Laasonen1993}
with a Nos\'e-Hoover thermostat~\cite{Nose1984, Hoover1985}. 
In order to match the 
experimental temperature~\cite{Hahlin2011} we thermalize the system at 
300~K for 3~ps, and then monitor the evolution of the energy levels for 6~ps. 
The calculated 
broadening is then included in the final PES spectrum by convolving the 
raw spectra with gaussians of the same width.
Quantum zero-point broadening could increase our calculated thermal broadening
by up to $\sim$0.1~eV~\cite{Cardona2005}, however calculating the precise
magnitude of this effect is beyond the scope of this work~\cite{Giustino20102}.
In order to account for configurational disorder we determine $\Delta_{\rm int}$
by repeating the DFT calculations for the six interface models with isolated
dyes described in Ref.~\citenum{Patrick2011}.
The escape depth of the photoelectrons is included by dividing the interface
models into layers coplanar to the surface and weighting the contributions 
from each layer using the factor ${\rm exp}(z/\lambda)$.
In this expression $z$ is the layer height and $\lambda$=5~\AA\ is the electron 
escape depth corresponding to 450~eV photons~\cite{Hahlin2011, Hufnerbook}.
For the surface coverage we take an areal density of 0.5 molecules/nm$^2$ based
on standard dye loading~\cite{Patrick2011}.

We now present our results.
As shown in Fig.~\ref{fig.2}(b) for a representative N3/TiO$_2$ interface
(model I2b of Ref.~\citenum{Patrick2011}), 
our requirement that the chromophore HOMO retains its gas-phase character 
is perfectly verified for the N3/TiO$_2$ interface. In fact the HOMO is well
separated from the underlying semiconductor ($>$5 \AA), and the overlap 
of the state shown in Fig.~\ref{fig.2}(b) with the HOMO state of the isolated 
molecule is 99\%. 

\begin{figure}
\includegraphics[width=\columnwidth]{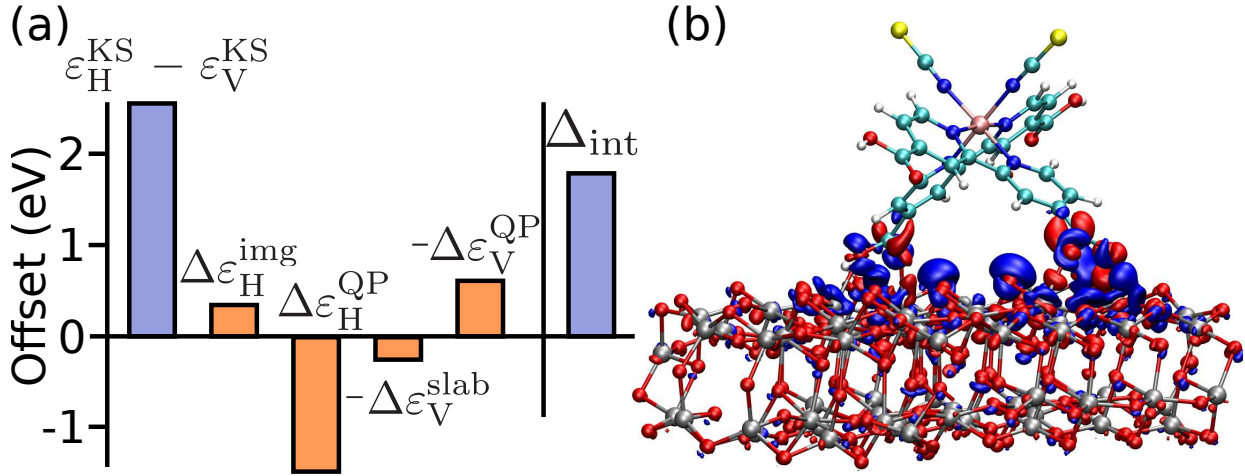}
\caption{\label{fig.3} (Color online)
(a) Bar chart indicating all the contributions to $\Delta_{\rm int}$
according to Eq.~(\ref{eq.2}). The bar $\Delta_{\rm int}$ on the right hand side
is obtained by adding up all the other bars.
(b) Atomistic model of the N3/TiO$_2$ interface (cf.~Fig.\ref{fig.2}), and
isodensity plot of the image charge induced by the removal of one electron from N3. 
The isodensity is $\pm0.0025$ \AA$^{-3}$ (blue/red).
The screening charge corresponds
to 0.20 electrons and is concentrated on the O atoms of the TiO$_2$ surface.
}
\end{figure}

Figure~\ref{fig.3} shows the calculated contributions to $\Delta_{\rm int}$
in Eq.~(\ref{eq.2}). For the model of Fig.~\ref{fig.2}(b) the KS energy-level 
alignment is $\e_{\rm H}^{\rm KS}-\e_{\rm V}^{\rm KS}$=2.60~eV,
while the quasiparticle energy offset is $\Delta_{\rm int}$=1.82~eV.
The largest correction to the KS alignments comes from the quasiparticle shift
of the N3 HOMO, $\Delta\e^{\rm QP}_{\rm H}=-1.50$~eV.
The quasiparticle shift of the TiO$_2$ VBT, $\Delta\e_{\rm V}^{\rm QP}=-0.62$~eV,
is rather large when compared to other semiconductors~\cite{Hybertsen1986}.
This result indicates that a simplified scissor correction 
would not be adequate for this interface.

An interesting result of this study concerns image-charge effects.
The redistribution of charge on creation of the photohole is illustrated in
Fig.~\ref{fig.3}(b). In addition to a screening charge layer concentrated  
on the TiO$_2$ surface and arising from image-charge effects~\cite{Neaton2006}, 
0.11 electrons/dye are transferred to the chromophore
through its anchor groups. This additional screening, which is not
observed in the case of physisorbed molecules~\cite{Neaton2006, Garcia2009},
is enabled by the presence of covalent bonds between N3 and TiO$_2$ which
provide a direct pathway for charge transfer. 
In order to capture this effect,
eventual GW calculations on entire interfaces will need to include 
self-energy corrections to the KS wavefunctions.
Our calculated correction $\Delta\e_{\rm H}^{\rm img}$=0.35~eV is smaller than the 
quasiparticle shifts but cannot be neglected in a quantitative analysis.

\begin{figure}
\includegraphics[width=0.7\columnwidth]{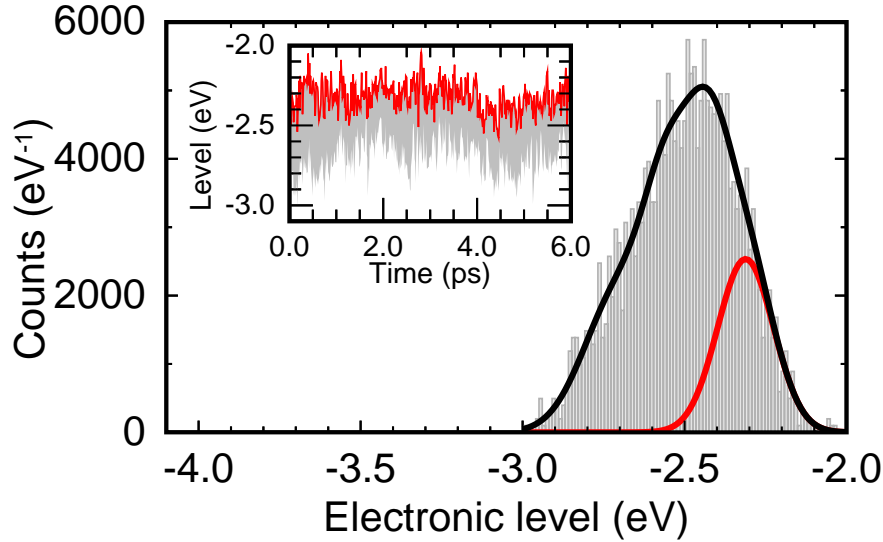}
\caption{\label{fig.4} (Color online)
Distribution of the four highest-energy KS eigenvalues of the N3/TiO$_2$ interface
model during a 6~ps molecular dynamics simulation at 300~K (histogram).
The eigenvalues are recorded at 0.01~ps intervals and the bin width is 0.01~eV.
The red (dark grey) line is a gaussian fit to the distribution of HOMO eigenvalues.
The black line is a fit to the eigenvalues distribution based on four gaussians.
Inset: time-evolution of the four highest-energy KS eigenvalues at 300~K.
The red (dark grey) line corresponds to the N3 HOMO, the (light) grey lines
correspond to the HOMO-1,HOMO-2, and HOMO-3 states.
}
\end{figure}

Figure~\ref{fig.4} shows the time-evolution at 300~K of the energy of the four highest 
electronic states at the N3/TiO$_2$ interface and the corresponding energy distribution. 
These states are similar in character to the dye HOMO, and consists of 
S-3$p$ and N-2$p$ orbitals from the thiocyanate ligands and of Ru-4$d$ 
orbitals~\cite{Fantacci2003}. In the ground state geometry these states 
lie within 0.3~eV of each other. The electron-vibration interaction 
is found to induce a thermal broadening
of 0.18~eV for each of these states. As a result the four highest-energy states
merge into a single peak, whose centre is redshifted by 0.19~eV with respect to
the HOMO in the ground state.

Configurational disorder is found to provide the largest source of spectral 
broadening.
In fact the KS HOMO/VBT separation varies between 1.81~eV and 2.46~eV
depending on the N3 adsorption mode. 
A similar trend was observed in a hybrid functional study of the related
N719 dye on a TiO$_2$ cluster~\cite{Angelis2007}. 

Figure~\ref{fig.1} reports the principal result of the present work, namely
the PES photocurrent at the N3/TiO$_2$ interface calculated using Eq.~(\ref{eq.1}). 
PES experiments reported HOMO/VBT offsets at the interface of
1.2~eV \cite{Hagfeldt2000}, 1.4~eV \cite{Rensmo1997, Westermark20023}, and
1.6~eV \cite{Schwanitz2007,Hahlin2011}. 
When we compare our calculations with the latest data from Ref.~\citenum{Hahlin2011}
[Fig.~\ref{fig.1}(inset)] it is clear that the DFT KS spectrum is in sharp disagreement
with experiment. 
Indeed standard DFT yields a finite density of states at the Fermi level,
due to the conduction band of TiO$_2$ being degenerate
with the HOMO of N3. The inclusion of Hubbard corrections lifts this
degeneracy but yields a HOMO/VBT offset which is $\sim$1~eV larger than
in the experiments.
On the other hand, the photocurrent calculated through Eqs.~(\ref{eq.1}),(\ref{eq.2}) 
by including quasiparticle shifts and spectral broadening are in
good agreement with the experimental PES data.
Given the huge complexity of the N3/TiO$_2$ interface and the first-principles
nature of our calculations we regard the comparison shown in Fig.~\ref{fig.1} 
as very satisfactory. We note in particular that the agreement between calculated 
and measured peak intensities is non trivial and 
is being reported here for the first time. Our results
indicate that the procedure described here captures the relevant 
physics of quasiparticle excitations at the N3/TiO$_2$ interface. 

A detailed analysis of the spectra in Fig.~\ref{fig.1} allows us to clarify some
aspects of PES at the N3/TiO$_2$ interface:
(i) Our calculations indicate that the thermal broadening is of the same magnitude 
as configurational disorder. This observation implies that the fingerprints
of specific dye adsorption geometries are blurred by temperature and cannot
be resolved in PES experiments at 300~K.
(ii) What is commonly identified as the N3 HOMO peak actually corresponds 
to an average over the four highest-energy states of the dye (Fig.~\ref{fig.4})
and is affected by a thermal redshift of $\sim$0.2~eV.
(iii) the finite PES intensity at binding energies around 3~eV in Fig.~\ref{fig.1}
does not arise from the TiO$_2$ substrate, but corresponds instead to lower-lying 
Ru-4$d$ orbitals.
(iv) The interaction with the semiconductor affects the quasiparticle
energies of N3 via image-charge screening (as for physisorbed molecules) 
and via charge transfer through the anchor groups.
Taken together these observations point to the necessity of calculating
{\it complete photoemission spectra} as opposed to individual KS energy levels 
in order to formulate a quantitative interpretation of measured PES data.

In conclusion, we developed a comprehensive theory of the quasiparticle energy-level alignment
and the photoemission spectra at complex chromophore/semiconductor interfaces. 
For the prototypical biomimetic solar cell interface between anatase TiO$_2$ and 
the N3 chromophore we were able to directly match the calculated photoemission spectrum
to experimental data and achieve quantitative accuracy.
The present work sets a new standard for reverse-engineering 
structure-property relations at solid/molecule interfaces
of direct interest for nanotechnology and energy applications.

This work is supported by the UK EPSRC and the ERC under the EU FP7 / ERC grant 
no. 239578.  Calculations were performed in part at the Oxford Supercomputing Centre. 
Figures rendered using VMD \cite{VMD}.

\end{document}